# Learned end-to-end high-resolution lensless fiber imaging toward intraoperative real-time cancer diagnosis

Jiachen Wu[1,2*], Tijue Wang[1], Ortrud Uckermann[3,4], Roberta Galli[5], Gabriele Schackert[3], Liangcai Cao[2], Jürgen Czarske[1,6,7,8*], and Robert Kuschmierz[1,6*]

[1]*Laboratory of Measurement and Sensor System Technique, TU Dresden, Helmholtzstrasse 18, 01069 Dresden, Germany*
[2]*State Key Laboratory of Precision Measurement Technology and Instruments, Department of Precision Instruments, Tsinghua University, Beijing 100084, China*
[3]*Department of Neurosurgery, University Hospital Carl Gustav Carus, TU Dresden, Dresden, Germany*
[4]*Division of Medical Biology, Department of Psychiatry, Faculty of Medicine and University Hospital Carl Gustav Carus, TU Dresden, Germany*
[5]*Department of Medical Physics and Biomedical Engineering, Faculty of Medicine Carl Gustav Carus, TU Dresden, Germany*
[6]*Competence Center BIOLAS, TU Dresden, Dresden, Germany*
[7]*Excellence Cluster Physics of Life, TU Dresden, Dresden, Germany*
[8]*Faculty of Physics, School of Science, TU Dresden, Dresden, Germany*
*\*E-mail: jiachen.wu@mailbox.tu-dresden.de; juergen.czarske@tu-dresden.de; robert.kuschmierz@tu-dresden.de*

**Endomicroscopy is indispensable for minimally invasive diagnostics in clinical practice. For optical keyhole monitoring of surgical interventions, high-resolution fiber endoscopic imaging is considered to be very promising, especially in combination with label-free imaging techniques to realize *in vivo* diagnosis. However, the inherent honeycomb-artifacts of coherent fiber bundles (CFB) reduce the resolution and limit the clinical applications. We propose an end-to-end lensless fiber imaging scheme toward intraoperative real-time cancer diagnosis. The framework includes resolution enhancement and classification networks that use single-shot fiber bundle images to provide both high-resolution images and tumor diagnosis result. The well-trained resolution enhancement network not only recovers high-resolution features beyond the physical limitations of CFB, but also helps improving tumor recognition rate. Especially for glioblastoma, the resolution enhancement network helps increasing the classification accuracy from 90.8% to 95.6%. The novel technique can enable histological real-time imaging through lensless fiber endoscopy and is promising for rapid and minimal-invasive intraoperative diagnosis in clinics.**

http://dx.doi.org/xxx

Early diagnosis of cancer is the key to improve the survival rate and cure rate of patients. Endoscopy plays an important role in the early stages of cancer diagnosis because many cancers can be observed *in situ* in body cavities and there is great benefit in guiding biopsy extraction by histopathological examination to confirm the diagnosis. The procedure of biopsy requires extracting tissues of the organs, fixation, sectioning, and staining for observation. Then, pathologists exercise judgment with microscopic images based on their knowledge and experience (Fig. 1a). Conventionally, It takes several hours to a few days for the surgeon to know the results of the diagnosis, which can reduce survival rates, especially for highly aggressive tumors. Moreover, improper resection can lead to additional risk, like internal bleeding or malfunctioning. To reduce the risk of surgical resection and save time for diagnosis, endoscopic realize real-time intraoperative diagnosis is required. The optical fiber is a key element for implementing multimodal microscopic approaches under *in vivo* conditions, then this endomicroscopy works as a minimally invasive probe applied at depth in living organisms to give instant diagnostic information.

Label-free nonlinear optical imaging techniques, providing a non-destructive approach for visualization of biomolecules, has proven to be a powerful tool for cancer research[1-6]. With the help of deep learning, these approaches can create virtual stained microscopic image bypassing the standard histochemical staining process[7-9]. In particular, nonlinear optical imaging modalities like coherent anti-Stokes Raman scattering (CARS), second harmonic generation (SHG), and two-photon excited auto-fluorescence (TPEF) microscopy have promising to complement conventional staining protocols[10-13]. To realize label-free endomicroscopy, the current approaches couple label-free scanning microscopes with standard-sized optical components and objectives into a rigid needle endoscope made from gradient index (GRIN) lenses[14, 15], or using fiber miniaturized resonant device to achieve fiber-scanning[16-18]. These approaches require distal optical elements such as scanners and lenses, such distal elements significantly enlarge the endoscope diameter, and increase tissue damage, which limits their clinical applications. Furthermore, the distal optics are complex, which increase costs and inhibit "single-use-probes" which are desired, since endoscopes are difficult to sterilize.

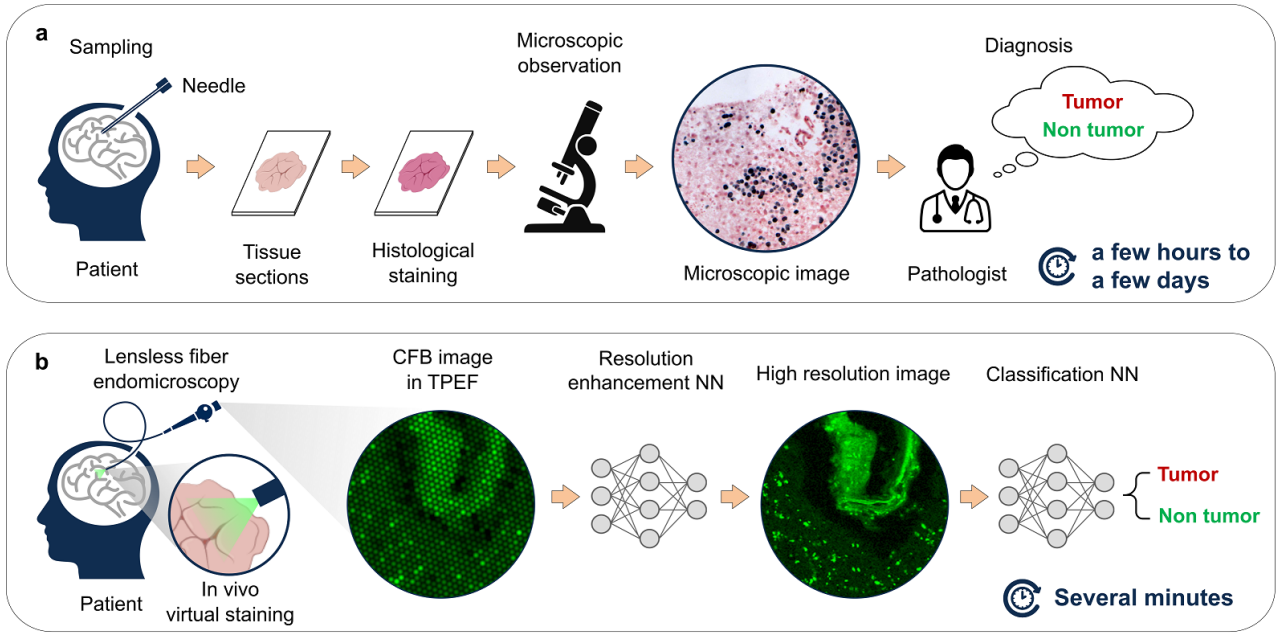

**Fig. 1 Workflow of biopsy diagnosis and end-to-end diagnosis. a**, A biopsy is the main way doctors diagnose most types of cancer diagnosis. It requires specialized surgical skills, histology laboratory and trained personnel. The whole procedure is cumbersome and time-consuming. **b**, Learning-based end-to-end diagnosis directly obtains the tissue image by endoscope, then using DNN to enhance resolution and give prediction of whether the tissue is tumor. It is promising for real-time and in situ diagnosis.

An alternative solution is utilizing coherent fiber bundle (CFB). CFB typically consists of thousands of cores, arranged in a honeycomb structure, with a common cladding. Each core acts as a pixel that can individually transmit intensity from the distal fiber end in the body to the observer at the proximal fiber end. Various optical techniques have been successfully integrated with CFB, which enriches the means of medical studies, such as light-field imaging[19], and holography[20-22], micro manipulation[23, 24]. Common CFB offers core diameters down to 2 μm and core-to-core distances down to 3 μm, which limits the spatial resolution and results in honeycomb artifacts. The honeycomb artifacts interfere with the identification of pathological tissue and increase the difficulty of diagnosis. Therefore, improving the resolution of CFB image is an urgent demand for identifying and diagnosing pathological tissues.

Conventional depixelation methods like Fourier domain filtering[25] and interpolation[26] can remove the honeycomb artifacts, but can not improve resolution. Optimization methods, such as maximum a posteriori estimation[27] and compressive sensing[28, 29], could improve the imaging quality by introducing prior information, but involve a time-consuming iterative procedure. With the multi-frame method, a sequence of images is captured with displacement or rotation to add information[30-32], however, the extra motion part and image registration increase complexity to the imaging system. Recently, it has been shown that deep learning offers nonlinear fitting abilities in image regression problems[33-37]. Thus, learning-based methods were applied to CFB for depixelation and resolution enhancement[38-41]. These works have limited sample types and numbers, however, which constrains the generalization capability of neural network.

In this paper, we proposed an end-to-end tumor diagnosis scheme using artificial intelligence (AI) technology to provide both high-resolution endoscopic image and tumors prediction result (Fig. 1b). Label-free images obtained directly by endomicroscopy are put in to a resolution enhancement network for CFB image, first. The resolution-enhanced network can remove the honeycomb-artifact of the CFB image and recovers high-resolution features beyond the physical limitations of the CFB. Then, the enhanced CFB image is fed into a classification network, which gives a prediction of tumor. Our approach adopts CFB to achieve single-shot imaging manner so that no scanning parts and post-processing algorithms required, which is an important advantage for intraoperative diagnosis. The instant diagnosis scheme dispenses the cumbersome process of biopsy, and is thus low-cost and patient-friendly.

To achieve this goal, a customized resolution enhanced network is used for training on both synthetic and real CFB images. A display-CFB-sensor imaging system is set up to collect arbitrary types of datasets. Furthermore, we demonstrate the positive impact of the resolution enhancement on neural network-based tissue classification. The results show the resolution enhancement network helps increasing the classification accuracy (prediction correct rate of all tissues) of glioblastoma from 90.8% to 95.6%. The proposed method can provide real-time and *in situ* diagnosis with endomicroscopy, which promotes the accuracy of targeted biopsies and the overall diagnostic yield.

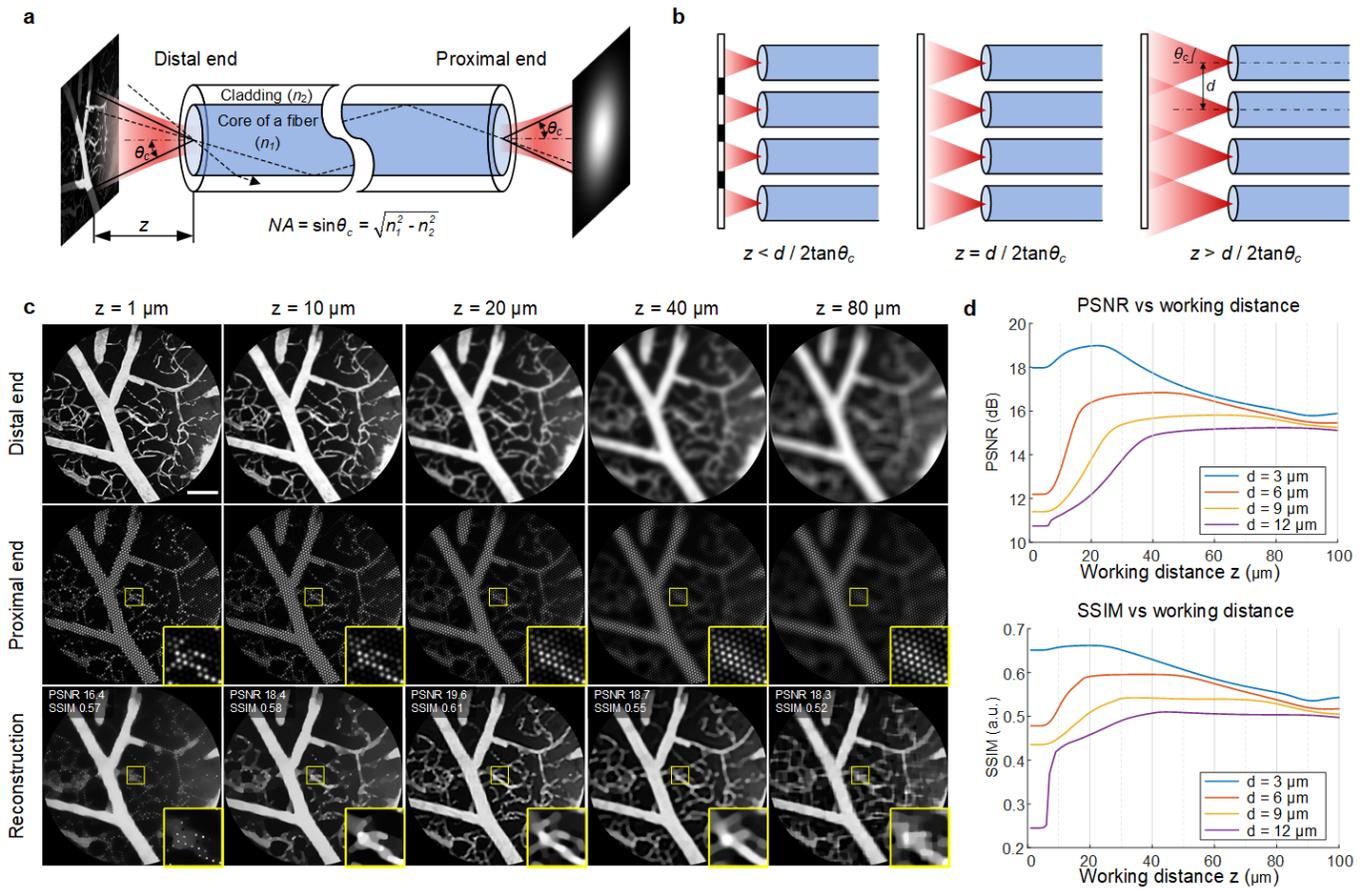

**Fig. 2 Optimal working distance for reconstruction. a**, Schematic of the transmission of an image by optical fiber as a single pixel. The working distance z and NA of fiber determine a cone space in which light can be coupled into fiber. **b**, the situation that CFB gather light from sample under different working distance z. **c**, The images at two ends of CFB and reconstructed images under different working distances. The reconstruction method is compressive sensing with TV regularization. Scale bar: 30 μm. **d**, Average PSNR and average SSIM at varying working distances over 10 images of mouse cortical vasculature.

## Results

**Optimal working distance for lensless endoscope.** Lensless endoscopes collect light directly from the distal CFB facet, making them tiny probes on the one hand, but making the light propagation complex on the other. Working distance is an important parameter for CFB imaging in lensless mode. The optimal working distance is related to the core spacing and numerical aperture (NA) of the fiber. The NA of a fiber is defined as the sine of the largest acceptable angle $\theta_c$ that an incident ray can have for total internal reflectance in the fiber core (Fig. 2a). The NA can be calculated according to the refractive indices of core and cladding. It determines a cone space in which light can be coupled into fiber. For a CFB with $NA = \sin\theta_c$, and core spacing is $d$, when a sample is very close to the fiber facet, that is $z < d / 2\tan\theta_c$, the region can be coupled into fiber cores does not completely cover the sample, which causes loss of information (Fig. 2b). Otherwise, when a sample is far away from the fiber facet, that is $z \gg d / 2\tan\theta_c$, causes a blurred observation (Fig. 2b). Therefore, choosing an optimal working distance is required to high-resolution image reconstruction.

Here we use the parameters of SUMITA HDIG to analyze the optimal working distance, where the core spacing is 3.0 μm and the acceptable angle $\theta_c = 8°$. A multiphoton microscopic image of mouse cortical vasculature[42] is chosen as true scene. We show the two ends of CFB and reconstructed images at five different working distances: 1 μm, 10 μm, 20 μm, 40 μm and 80 μm (Fig. 2c). Reconstruction is performed using compressive sensing with total variation regularization. When the working distance is 1 μm, the true values of the region between the cores, called dead-space, cannot be recovered. As the working distance increases, the information that was originally in dead-space can be collected by the adjacent cores, so that the recovery become possible. However, as the distance increases, the image at distal facet is blurred; this in turn makes it more difficult to recovery details. In this case, 20 μm is the best working distance.

To disclose the relationship between optimal working distance and core spacing, we use an optimization algorithm to calculate the reconstructed image quality under different working distance and different core spacing to find the optimal working distance. Peak signal to noise ratio (PSNR) and structural similarity index measure (SSIM) are adopted for quantitative evaluation of image quality. The results show reconstruction can achieve optimum quality when the working distance is in the range of $2d \sim 8d$ (Fig. 2d).

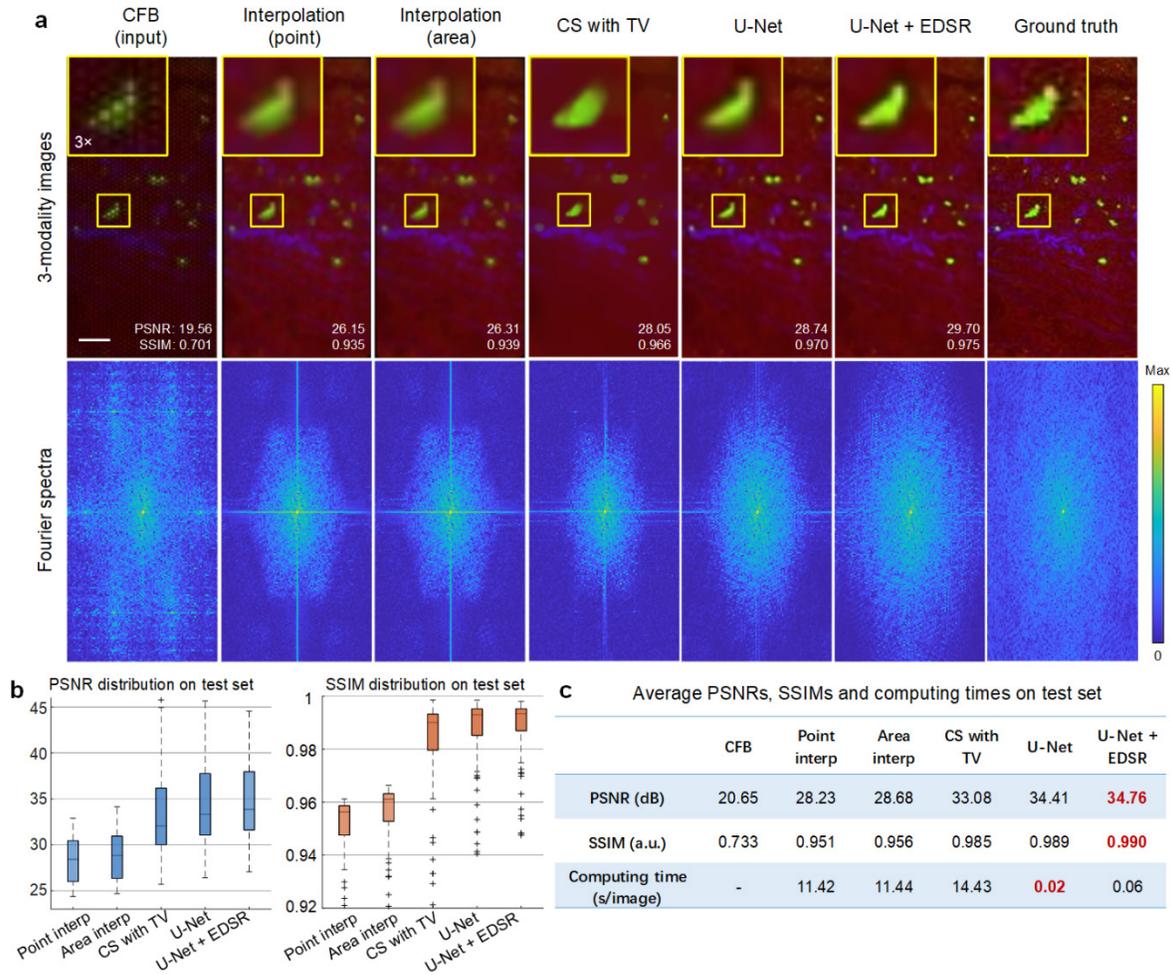

**Fig. 3 Image enhancement for the synthetic CFB images. a**, Comparison of enhanced images with ground truth by different methods in spatial and frequency domain. Scale bar: 20 μm. **b**, Quantitative evaluation of the image quality of different methods in terms of PSNR and SSIM. Box plot: center line, median; box limits, upper and lower quartiles; whiskers, 1.5× interquartile range; plus sign, outliers. **c**, Average PSNRs, SSIMs and computing times of different methods. The U-Net + EDSR provides best image quality and fast computation.

**Resolution enhancement model based on U-Net + EDSR.** A simple way to generate datasets is to synthesize CFB images from ground truth (GT) images. Here we adopt label-free multiphoton images as GT to synthesize the CFB images. The images were obtained using a multi-modal microscope, image modalities are CARS, TPEF, and SHG. Originally the images were used for tissue classification using linear discriminant analysis (LDA)[43]. The three modalities are combined into single RGB image. The lateral resolution is 1 μm. Image size is 208 × 104 pixels. Then CFB imaging model is applied to generate pixelated images. The optic fiber model of SUMITA HDIG is simulated for dataset generation, where the core diameter is 2.0 μm and core spacing is 3.0 μm.

The well-trained network was applied to the test set, which contains 200 RGB images of 9 tumor and 2 non-tumor types. The comparison of the PSNR and SSIM distribution using the point interpolation, area interpolation, compressive sensing (CS) with total variation (TV) regularization, U-Net and U-Net + EDSR is demonstrated (Fig. 3a). The instance shows that the learning-based methods are superior to all others. Moreover, the U-Net + EDSR configuration shows edges more clearly than U-Net-only configuration. The U-Net architecture could learn the features at different resolution scales, but U-Net lacks deeper layers to learn complex and variable features in each scale. EDSR consists of deep residual blocks, so connecting EDSR with U-Net can make up for the limited ability of network characterization at high resolution scale. The enhanced image by U-Net + EDSR has prominent target features, which can help a doctor discriminate tissue type intraoperatively. By quantitatively analyzing the quality of reconstruction on 200 images in test set, U-Net + EDSR has the highest average PSNR and SSIM, and has more centralized distributions than U-Net (Fig. 3b and c). The computing time is much faster than interpolation and CS methods (Fig. 3c).

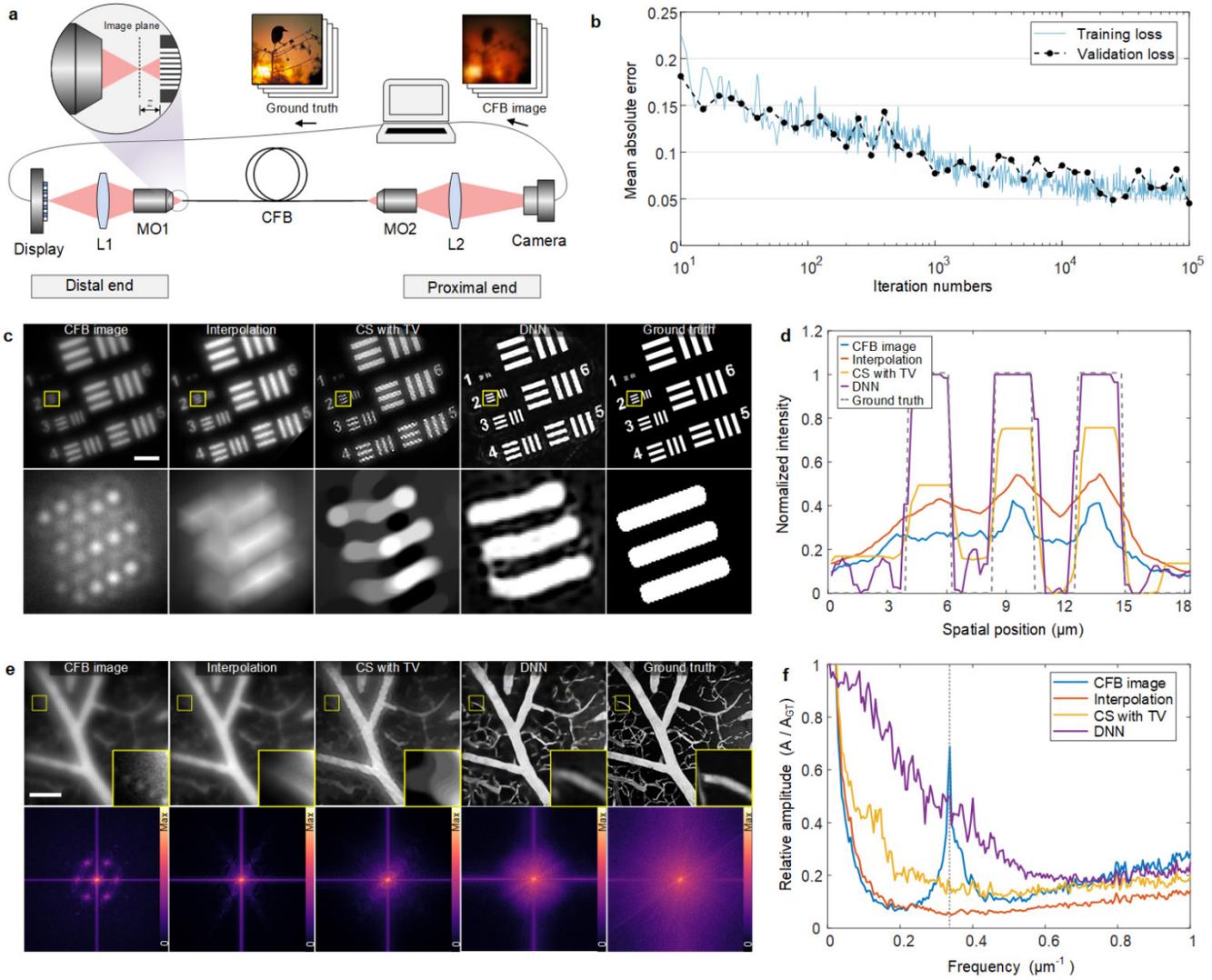

**Fig. 4 Image enhancement for the real CFB images. a** Experimental setup for dataset acquisition. The displayed images are projected to the optimal working distance z = 20 μm in front of the fiber facet. L, lens; MO, microscopic objective. **b** The change of loss function during training process. **c** Comparison of imaging results between different methods for resolution chart. Scale bar, 50 μm. **d** Cross section on Group-2 for each method. **e** Comparison of frequency-domain characteristics of different image enhancement methods. The closeups show the details of tiny vessels. Scale bar, 40 μm. The corresponding spectrum maps are shown in the bottom of spatial images. **f** Radial normalized amplitude of spectrums. The dashed line indicates the core sampling frequency.

**Resolution enhancement for experimentally acquired images.** Defects in real CFBs make the actual images deviate from simulations. For example, irregular core shape and nonuniform refractive index may cause the inner-core coupling or excite cladding modes. These factors can lead to failure of network prediction for real CFB images. To address the problem, a display-CFB-sensor imaging system is setup to obtain pairs of real CFB images and GT images directly (Fig. 4a). The GT images are displayed on a mobile phone screen (Samsung S10) with pixel pitch 46.1 μm. The screen is projected onto the CFB (Sumita HDIG) facet using a 40× objective and tube lens. The CFB has around 10,000 fiber cores and a core pitch of 3 μm. The GT images are projected 20 μm away from the facet in accordance to the results in methods section. The distal CFB facet is then imaged onto a camera with 2.2 pixel pitch. The magnification is adjusted to ×2.7 so that the CFB occupy the same pixels number with GT image. We adopt 5,500 natural images from ImageNet [44] as ground truth to display on the screen. All images are scaled to 512 × 512 pixels for display. We train the network for $10^5$ iterations, and the validation loss always remains the same level with training loss (Fig. 4b), which shows the network has good generalization performance.

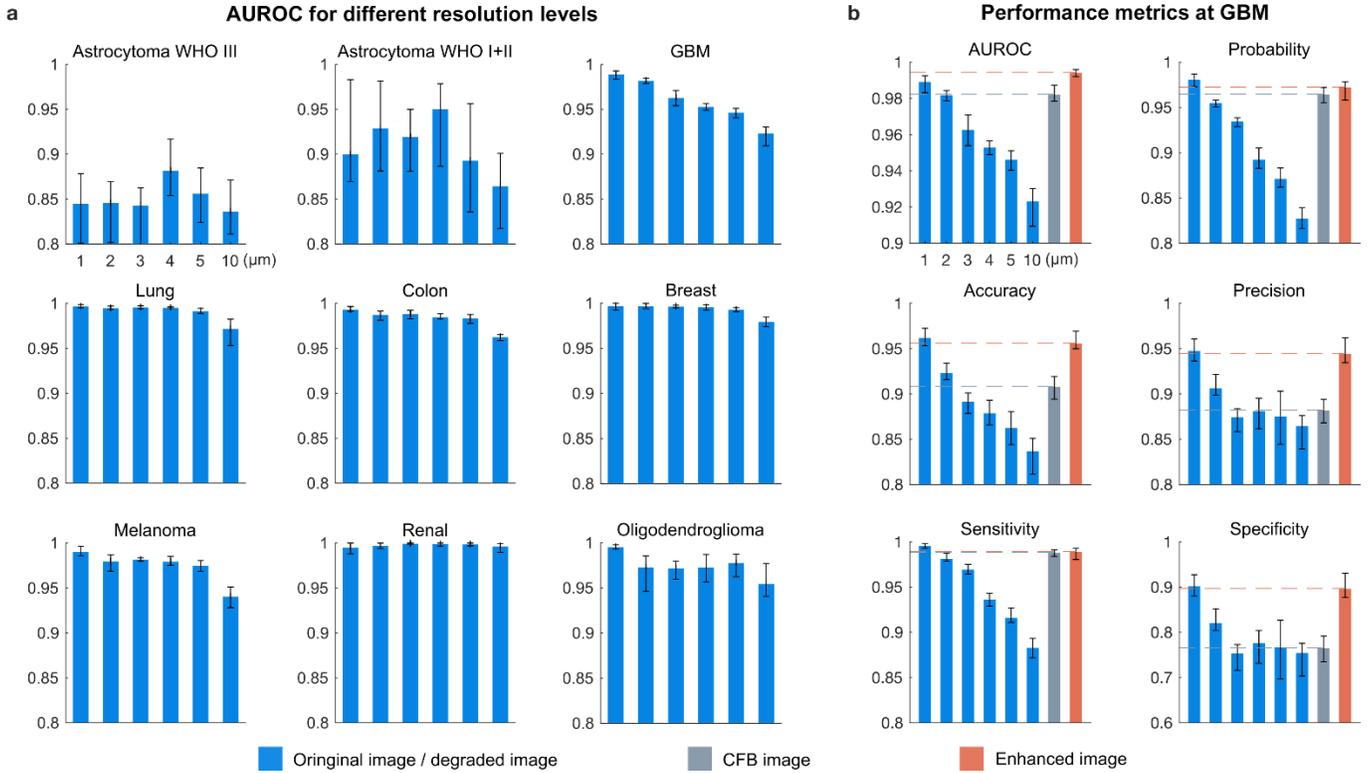

**Fig. 5 Comparison of classification performance for different image types. a** AUROC values of the classification networks trained on different resolution levels, **b** The classification performance for GBM. The classification networks are individually trained and predicted for each data type CFB images and resolution enhanced images. All the results are based on TPEF modality.

A customized resolution chart is displayed on the screen to experimentally test the imaging resolution and contrast. The group number indicates the line width in pixels. According to the pixel pitch of display and objective magnification, the minimum line width in projected image is 1.15 µm. For all the reconstruction methods, Group-2 can be resolved (Fig. 4c), thus an upper bound for the maximum resolution is 2.3 µm or 217 lp/mm. This is better than the core-to-core distance of 3 µm and achieved by the increased working distance. While the increased working distance decreases the contrast initially, the EDSR enables contrast enhancement again. The cross sections of Group-2 show the imaging contrast (Fig. 4d). Defining the contrast as $(I_{max} - I_{min}) / (I_{max} + I_{min}) \times 100\%$, where the $I_{max}$ and $I_{min}$ represent the average intensity of the white and black regions respectively. Then the contrast of the CFB image, interpolation, CS and DNN methods are 65.6%, 72.1%, 75.2%, and 86.3% respectively.

Furthermore, the frequency-domain characteristics of reconstructed images are analyzed. The validation on mouse vasculature image shows that the DNN method can recover the most high-frequency components (Fig. 4e). To explicitly compare the frequency component that can be recovered by various methods, the amplitude spectra are averaged along the radial coordinate first, and normalized to the spectral amplitude of the ground truth. The results are plotted in frequency-amplitude-curves (Fig. 4f). Note that the curve peak of the CFB image indicates the sampling frequency of the fiber core pitch at 0.33 µm$^{-1}$, which is illustrated by the dashed line. According to Nyquist-Shannon sampling theorem, spectrum aliasing occurs when the signal frequencies exceed half of the sampling frequency. Interpolation methods only flatten the frequency curve but do not introduce high frequency components. The CS method introduces prior information through TV regularization, and high frequency components are slightly improved. However, TV regularization with a scalar weight is based on spatially invariant assumptions, which make it difficult to handle both homogeneous features and regions with rich details. In contrast, the DNN method can learn various features from the dataset, so that it can adaptively restore variable image features.

**Influence of image resolution on tumor classification results.** Fluorescence imaging for tissue provides rich information for tumor diagnosis. A high-resolution fluorescence image can indicate tumor margin[45], degree of tumor progression[46], and other pathological features for fine-grained analysis[47]. However, for the tumor delineation, only binary discrimination of tumor and non-tumor is necessary. If the classification results are not sensitive to resolution, a low resolution and large field of view endoscope can be adopted for rapid tumor screening. Otherwise, high resolution imaging technology is required.

To test the influence of image resolution on tumor classification results, we applied Gaussian filters with different kernel sizes to reduce the resolution for the TPEF images of biopsies of human brain towards malignant and benign tumors. The full width at half maximum of Gaussian filter is used to represent the resolution of degraded images. A Visual Geometry Group-19 (VGG-19)[48] classification network is trained on each resolution level. Here we test the classification performance and show the area under receiver operator characteristic curve (AUROC) on the resolution of 1 μm (original resolution), 2 μm, 3 μm, 4 μm, 5 μm and 10 μm (Fig. 5a). The AUROC is used for each of test datasets as the performance metric. For each case, the training process is repeated for 5 times with different patients randomly chosen for training. The upper and lower limits of error bar indicate the maximum and minimum achieved values.

For astrocytoma WHO I/II and III, the AUROCs show low correlation with resolution and overall low performance. Actually, the problem of overfitting appeared in the training process of these two tumor types (Supplementary Fig. 1). The reason is perhaps great feature difference between training set and test set. More extensive patient samples could address this problem.

For the brain metastases of lung cancer, colon cancer, renal cancer, breast cancer and of malignant melanoma, the AUROCs keep an almost constant and high level which are all above 0.96 for 5 μm resolution and above. The results demonstrate the VGG-19 network can be trained and tested on CFB images (3 μm core spacing in our experiment) without resolution enhancement for these tumors.

For anaplastic oligodendroglioma WHO III, only at the 1 μm resolution, AUROC is stable and has a high level. This tumor type may have more high-frequency features (Supplementary Fig. 2) so that the diagnosis strongly dependents on high resolution images.

For glioblastoma (GBM), the AUROCs show a nearly linear decrease with resolution. This is a great property for the CFB imaging since the resolution enhancement technique could play a role in improving the classification performance. Thus, it can be used to investigate the performance of the proposed resolution enhancement network.

**Efficient improving GBM classification performance using resolution enhancement network.** GBM is a highly aggressive type of brain tumor, so early diagnosis and treatment are of great significance for prolonging the life span of patients. Here we use six metrics of AUROC, probability, accuracy, precision, sensitivity and specificity to evaluate the effect of using resolution enhancement network on classification performance. We individually trained networks on microscopic images, CFB images and resolution enhanced images, and the classification results on different resolution are shown for comparison (Fig. 5b). In all the metrics, the resolution enhanced images have better performance compared to the CFB images. This verifies the proposed resolution enhancement network can efficiently improving GBM classification performance. The average accuracy of microscopic images, CFB image and enhanced image are 96.2%, 90.8%, 95.6%, respectively. In this case, the honeycomb-artifacts of CFB images deteriorate some characteristics of tumor morphochemistry and reduce the classification performance slightly. Then the resolution enhancement process can reconstruct features and improve classification accuracy to the same level as the microscopic image.

## Discussion

The expected clinical application of the proposed high-resolution endoscopic imaging is the instant tumor recognition, namely predicting the probability of a potential cancer patient rapidly and automatically by analyzing multiple images of the suspected tissue without performing time-consuming biopsies, and reduce the additional risk, such as internal bleeding or malfunctioning.

However, single image-based resolution enhancement is an ill-posed problem because one low-resolution image can yield several possible high-resolution images. The deep learning method is to learn the prior information to find the most likely solution. It is crucial for biomedical imaging to see if the resolution improvements brought by deep learning are real and effective. In computer vision, the super-resolution network is a feature-oriented model, which learns the feature mapping from low resolution to high resolution. If a new feature appears in the original image but does not exist in the training set, the model tends to use existing features to predict new features. It may generate visually satisfying images; however, the fake features may lead to misprediction in clinical application.

Our learning-based resolution enhancement approach includes a fiber bundle image acquisition system to obtain training data, thus the dataset involves the physical model of image degradation. Then the networks can learn inverse physical models rather than just feature mapping. Even the network is trained on natural image, it is still suitable for medical image enhancement. In addition, experimentally capturing training data allows for incorporating all the non-idealities present in a real system into the dataset, thereby making the method generalize for various real clinical scenarios.

The reliability of uninterpretable models is the key concern when incorporating models into clinic practice. Since the explainable AI has made some progress in real-world applications[49,50], the future research will focus on the mechanisms of tumor classification by DNN, and improve the network to provide more robust and abundant information, such as the degree of the lesion and lesion area, so that deep learning methods can provide trusty means of medical diagnoses.

## Conclusion

In conclusion, we present an end-to-end endoscopic diagnostic protocol for intraoperative real-time cancer diagnosis. Firstly, high-resolution imaging through a fiber-based endoscope is achieved by resolution enhancement network. The proposed U-Net + EDSR network can learn both the CFB imaging model and the image prior, so that the inherent honeycomb-like artifact can be eliminated and high frequency information could be retrieved. Then a VGG-19 classification network is trained on TPEF images to provide AI-assisted diagnosis. The resolution enhancement results show the average PSNR can be improved by more than 10 dB, and the contrast can be improved by 20%. The proposed resolution enhancement network can further improve the classifier performance for GBM, where classification accuracy increased from 90.8% to 95.6%. The proposed endoscopic imaging pipeline can provide better guidance to surgeons with respect to identification and localization of the tumor during surgery.

## Methods

**CFB-based imaging model.** The CFB translates the intensity distribution on the distal facet to the proximal facet in a pixelated manner. Considering a sample placed at a distance $z$ from the distal facet, each fiber core couples the light within its acceptance angle, implementing a weighted sum of the original image. Then the image degradation can be modeled as convolution with a point spread function (PSF). The PSF could be contributed by three parts: distance attenuation term, source divergence angle and facet coupling efficiency. The distance attenuation term follows inverse-square law, and considering the fiber critical angle, rays with an incident angle greater than the critical angle $\theta_c$ cannot be coupled into the fiber. The distance attenuation term could be formalized as:

$$I_{dist}(\vec{r};z) = \begin{cases} \dfrac{z^2}{z^2+|\vec{r}|^2}, & \theta \leq \theta_c \\ 0, & \theta > \theta_c \end{cases}, \quad (1)$$

where $\theta = \arctan(|\vec{r}|/z)$. The facet coupling efficiency depends on collection aperture of the fiber. An approximation model for the facet coupling efficiency is a Gaussian distribution, which can be parameterized as follow[51]:

$$I_\sigma(\vec{r};\sigma) = \exp\left[-|\vec{r}|^2/2\sigma^2\right] \quad (2)$$

where $\sigma$ denotes the width of the Gaussian function. When the full width at half maximum (FWHM) of Gaussian distribution is equal to the fiber core diameter $d$, $\sigma$ has the value $d/2\sqrt{2\ln(2)}$. Assuming the light source has uniform distribution in all angle, the total PSF could be modelled as:

$$\mathrm{PSF}(\vec{r};z,\sigma) = I_\sigma(\vec{r};\sigma) * I_{dist}(\vec{r};z). \quad (3)$$

Then each fiber cores samples the intensity at distal facet:

$$Y_i = \left[\mathrm{PSF}(\vec{r};z,\sigma) * X(\vec{r})\right] \cdot \delta(\vec{r}-r_i) \quad (4)$$

where $r_i$ is the position vector of the core centers. $Y_i$ represents the $i$th downsampling measurement for the high-resolution image $X$. Then the fiber bundle conveys the sampled intensities to the proximal end. For single mode cores, only the $\mathrm{LP}_{01}$ mode can be transmitted over optical fibers so that all the cores have the same relative intensity distribution at the proximal facet. The $\mathrm{LP}_{01}$ mode of optical fiber is often expressed approximately by the Gaussian field[52]. Then applying the convolution again to form the observed honeycomb-like image:

$$I_{honeycomb}(\vec{r}) = \frac{1}{2\pi\omega^2}\exp\left[-|\vec{r}|^2/2\omega^2\right] * \sum_i^N Y_i(\vec{r}) \quad (5)$$

where deviation $\omega$ is the equivalent mode field radius, which is related to the fiber parameters.

**Compressive reconstruction.** The true resolution of the honeycomb-like image is the number of fiber cores, which is usually much less than the number of sensor pixels. To reconstruct the high-resolution image from the CFB image is an ill-posed problem. Compressive sensing (CS) is a typical method to deal with such problems. Compressive sensing is a powerful signal reconstruction framework and provides complete theoretical support for image reconstruction. It states a given signal could be accurately reconstructed with fewer samples or measurements, which is not necessary to satisfy the Nyquist's sampling theorem. CS theory indicates the conditions for accurate reconstruction is sparsity and incoherence. The sparsity means there are many zero-valued elements in the signal itself or in some transform domain. The incoherence means the sensing matrix and representation matrix are uncorrelated. For natural images, it is sufficiently sparse with its representation in the gradient domain or wavelet domain. In CFB imaging, the general image degradation can be expressed as convolution form. The convolution matrix is the Toeplitz matrix, which will satisfy the constrained isometric property with high probability[53]. Therefore, CFB imaging can approximately satisfy the compressed sensing condition and achieve high-quality reconstruction.

Since the valid measurements for CFB imaging is the core intensities, the CFB image can be represented by integrating each intensity in the core and rearrange them into 1D vector $\vec{Y}$. The original image can be represented in 1D vector $\vec{X}$ in lexicographical order. The convolution with the PSF can be represented as Toeplitz matrix $W$. Then Eq. (4) can be rewritten in matrix-vector form:

$$\vec{Y} = DC\vec{X} + \vec{E}, \quad (6)$$

where $\vec{E}$ is the additive noise. $D$ is the downsampling operator. All linear manipulation can be simplified into CS literature form to

$$\vec{Y} = W\vec{X} + \vec{E}, \quad (7)$$

where $W$ represent combines the convolution and downsampling operation in one operation. Assuming the fiber bundle has $M$ cores and the camera has $N$ pixels, where $M < N$. Then $\vec{Y}$ is the $M \times 1$ vectorized observation, $\vec{X}$ is the $N \times 1$ vectorized object, and $W$ is a $M \times N$ matrix. Obviously, this is an underdetermined system. The traditional solution for this problem is the least squares method, which is to solve the following optimization problem:

$$\hat{X} = \arg\min_{\vec{X}} \|W\vec{X} - \vec{Y}\|_2^2. \quad (8)$$

However, additive noise in measurements greatly affects the accuracy of the results. Thus, it is necessary to introduce a regularization term to stabilize the solution. Then the optimization problem becomes:

$$\hat{X} = \arg\min_{\vec{X}} \|W\vec{X} - \vec{Y}\|_2^2 + \tau\Phi(\vec{X}) \quad (9)$$

where $\tau$ is a coefficient that balances the regularization term and the data fitting term and $\Phi$ is the regularization term representing the prior. Reconstructions were performed using the two-step iterative shrinkage/threshold (TwIST) algorithm[54,55] with total variation (TV) regularization.

**Data acquisition.** Our analysis based on the multiphoton image set comes from Uckermann's paper[43]. It includes Coherent anti-Stokes Raman Scattering (CARS), Two-Photon Excited Fluorescence (TPEF), and Second Harmonic Generation (SHG) microscopy images on cryosections of brain tumors of 382 patients and 28 human nontumor brain samples. The previous research verified the combined analysis of texture parameters of the CARS and TPEF signal is most suited for the discrimination of nontumor brain versus brain tumors. The classification includes different tumor types (low- and high-grade astrocytoma, oligodendroglioma, glioblastoma, recurrent glioblastoma, brain metastases of lung, colon, renal, and breast cancer and of malignant melanoma), and demonstrate a correct rate of 96% (sensitivity: 96%, specificity: 100%) by using linear discriminant analysis (LDA) method. LDA is a dimensionality reduction technique, which project the features in higher dimensional space onto a lower-dimensional space that facilitate the classification. Before applying LDA, texture analysis methods were used to extract statistics features to characterize decisive information.

We reproduce the classification using deep neural network (DNN) in the case of single modality: TPEF, two modalities: CARS + TPEF, and three modalities: CARS + TPEF + SHG. TPEF is chosen in single modality for comparison because of its high classification accuracy[43] and straightforward implementation in a fiber probe[56]. In the tumor classification task, it is necessary to divide the dataset by patients. Specifically, for the dataset of each tumor type, the existing patients were randomly divided into training, validation, and test set, respectively. Then the tumor images corresponding to the patients were assigned to each set, whereby it can be ensured that what the network learns is the features of the tumor type, rather than the patient-specific tissue features. In our case, the total number of patients used for training, validation and testing are 311, 33, and 37, respectively (see the patient distributions of each tumor type in Supplementary Table 1). For each type of tumor, we randomly assigned 2 patients with non-tumor to participate in classification training. We use accuracy, sensitivity (correct rate of tumor), specificity (correct rate of non-tumor) to evaluate the overall classification performance. The results show the single modality has a correct rate of 98.2% (sensitivity: 97.3%, specificity: 100%), which has almost equivalent performance with multi-modalities (Supplementary Table 2). DNN method has similar performance to LDA for combined CARS + TPEF modalities, and is superior than LDA for single TPEF modality. The results verify the feasibility of clinical exploitation using two-photon fluorescence endomicroscopy systems.

**Network architecture and training process.** Recently deep learning for single image super-resolution have been employed and achieved the superior performance. Here we employ U-Net[57] with enhanced deep super-resolution (EDSR) networks[58] for CFB imaging to achieve both a fast and a high-quality reconstruction (Supplementary Fig. 3). The U-Net part consist of a series of down sampling and up sampling blocks to learn the features at different resolution scales. We remove the batch normalization layers in both networks, since they get rid of range flexibility from networks by normalizing the features. For the EDSR part, the network is mainly composed of residual blocks in series. The additional scaling layer in the residual block of the EDSR helps to stabilize the training progress when a large number of filters is used. A convolution layer is used to extract features at the beginning and the end of all the residual blocks, respectively. A skip connection connects these two convolution layers. Finally, the image is output through a convolution layer. All convolutional layers use filters of size 3 × 3. Since there is no need to increase the image size in our case, we remove the upsample layer in original model to keep the output has the same size with the input. The depth (the number of residual blocks) is 32 and feature number is 256.

The loss function is evaluated on both pixel-wise and feature-wise metrics. The mean absolute error is calculated as pixel-wise metric. The pretrained VGG-16 is used to define the perceptual loss function that measures perceptual differences in output and ground truth label. The total loss function is the sum of these two terms with an adjustable weighting coefficient.

We trained two networks for synthetic images and real images respectively. For the synthetic images, 10,000 images from multiphoton biopsies images of human brain tumors are randomly selected and then cropped to the size of 192 × 96. 9,500 of them are used for the training, 300 for the validation and 200 for the testing. For real images, the minibatch size is 4. The learning rate is initialized to $10^{-4}$ for all layers and decreases by a factor of 0.5 for $2 \times 10^3$ iterations for a total of $10^4$ iterations. The training was run on a workstation with 32 AMD Ryzen 9 3950X 16-Core Processors and a NVIDIA RTX A6000 GPU.

**Acknowledgement**
We thank the German Science Foundation (DFG) for extensive funding of the project and Edmund Koch, TU Dresden for support.

**Funding**
German Research Foundation (DFG) (CZ55/47-1) and EKFZ (BrainAce), Tsinghua Scholarship for Overseas Graduate Studies (2020023).

**Conflict of interest**
The authors declare no conflict of interest.

**Author contributions**
J.W., T.W. and R.K. designed and conducted the experiments and programmed the algorithm. O.U., R.G. and G.S. acquired and analyzed the biological data. J.W. and T.W. wrote the manuscript. R.K. and J.C. conducted the project management. All authors contributed to the preparation of the paper.